\documentclass[12pt, a4paper]{article}

\pdfoutput=1
\usepackage{hyperref} 
                                                                 
\usepackage{amsmath}
\usepackage{amssymb}
\usepackage{multirow}
\usepackage{arydshln}

\usepackage{graphicx}
\usepackage{inputenc}
\usepackage{xcolor}
\usepackage{tikz}
\usetikzlibrary{matrix}
\usetikzlibrary{calc,fit}
\usepackage{slashed}

\newcommand{\be}{\begin{equation}}
\newcommand{\ee}{\end{equation}}
\newcommand{\bea}{\begin{eqnarray}}
\newcommand{\eea}{\end{eqnarray}}
\newcommand{\eq}[1]{Eq.~(\ref{#1})}

\def\bma#1{\mbox{\boldmath{$#1$}}}

 \def\tq{\tilde{q}}
 
  \def\dal{\dot\alpha}

    \def\bsig{\bar\sigma}

\def\lra#1{\overset{\text{\scriptsize$\leftrightarrow$}}{#1}}
\def\bma#1{\mbox{\boldmath{$#1$}}}

\def\lra#1{\overset{\text{\scriptsize$\leftrightarrow$}}{#1}}

\begin{document}

\thispagestyle{empty}
\begin{titlepage}

\today
\vspace{.3in}

\vspace{1cm}
\begin{center}
{\Large\bf\color{black} 
One-loop non-renormalization results in EFTs}\\
\bigskip\color{black}
\vspace{1cm}{
{\large  J.~Elias-Mir\'o$^{a,b}$, J.R.~Espinosa$^{a,c}$, A.~Pomarol$^{b}$}
\vspace{0.3cm}
} \\[7mm]
{\it {$^a$\, IFAE, Universitat Aut{\`o}noma de Barcelona,
   08193~Bellaterra,~Barcelona}}\\
    {\it {$^b$\, Dept.~de~F\'isica, Universitat Aut{\`o}noma de Barcelona, 08193~Bellaterra,~Barcelona}}\\
{\it $^c$ {ICREA, Instituci\'o Catalana de Recerca i Estudis Avan\c{c}ats, Barcelona, Spain}}
\end{center}
\bigskip

\begin{abstract}
 In Effective Field Theories (EFTs) with higher-dimensional operators 
 many  anomalous dimensions vanish at the one-loop level.
With the use of supersymmetry, and a classification of the operators according to
their embedding in super-operators,  we are able  to understand  why  many of these anomalous dimensions are zero.
The key observation  is that  
one-loop contributions from  superpartners  trivially 
vanish in many cases under consideration, making the superfield formalism a powerful tool even for
non-supersymmetric models.
We show this in detail in a simple $U(1)$ model with a scalar and fermions, and explain how to extend this to  SM EFTs
and  the QCD Chiral Langrangian.
This  provides an    understanding of  why most "current-current"  operators do not 
renormalize  "loop" operators at the one-loop level,  and allows  to  find the  few exceptions
 to this ubiquitous rule.

\end{abstract}
\bigskip

\end{titlepage}

\section{Introduction}

Quantum Effective Field Theories (EFTs)  provide   an excellent framework 
to describe   physical systems, most prominently in  particle physics, cosmology and condensed matter. 
With the recent discovery of the Higgs boson and the completion of  the SM, EFTs  have  provided a systematic approach to smartly parametrize our ignorance on possible new degrees of freedom at the TeV scale. Any theory beyond the SM, with   new heavy degrees of freedom, can be matched into an EFT that consists of operators built out solely with the SM degrees of freedom.

Recently, there has been much effort   put into  the determination of  the one-loop anomalous dimensions of the 
dimension-six   operators of the SM EFT~\cite{GJMT,EEMP1,EEMP2,Jenkins:2013zja,EGGM}. 
This has  revealed  a rather intriguing structure  in the anomalous-dimension matrix, with plenty of vanishing entries that are a priori allowed by all symmetries.
 Some vanishing entries are trivial since no possible diagram exist. Nevertheless, some  of them 
show intricate cancelations without any apparent reason.
Similar cancelations had been observed before in other EFTs
(see for example~\cite{Gasser:1983yg, Grinstein:1990tj}).

To make manifest the pattern of zeros in the matrix of anomalous dimensions, 
it is crucial to work in the proper basis.
Refs.~\cite{EEMP1,EEMP2} pointed out the importance of working in   bases with operators classified 
as "current-current" operators and  "loop" operators.
The first ones, which we call from now on $JJ$-operators, were defined to be those operators that can be generated 
as a product of spin-zero, spin-1/2 or spin-one currents of renormalizable theories \cite{SILH,Low:2009di,EEMP2}, while the rest
 were called "loop" operators.~\footnote{This classification is well-defined regardless of the specific UV-completion.  Field redefinitions (or use of the equations of motion)  do not mix $JJ$-operators and loop-operators.} 
In this basis it was possible to show  \cite{EEMP1} that 
some class of loop-operators were not renormalized by $JJ$-operators, suggesting a kind of generic non-renormalization rule.
The  complete pattern of  zeros in the SM EFT was recently provided  in
  Ref.~\cite{Alonso:2014rga}  in  the basis of 
\cite{Grzadkowski:2010es}, a basis that also  maintains the   separation between $JJ$- and loop-operators.
A classification of  operators based on  holomorphy  was suggested  to be a key ingredient
 to understand the structure of zeros  of  the anomalous-dimension matrix~\cite{Alonso:2014rga}.

In the present paper we provide  an approach to understand  in a simple way the vanishing of  anomalous-dimensions.
The reason behind many cancelations is the different Lorentz structure of the operators that  makes it  
impossible to mix them at the one-loop level. Although it is possible to show this in certain cases by simple inspection of the one-loop diagrams, we  present a more  compact and systematic approach based on the superfield formalism.
For this reason  we  embed  the EFT  into an  effective   superfield theory (ESFT),
and classify the  operators  depending on their  embedding into super-operators.
Using the ESFT, we are able to show by a simple spurion analysis 
(the one used to prove  non-renormalization theorems in supersymmetric theories)
 the absence, in certain cases, of mixing between operators of different classes.
We then make the important observation that the superpartner contributions to the one-loop renormalization under consideration trivially vanish in many cases. 
  This   allows us to conclude that some of the non-renormalization
  results of the ESFTs apply to the non-supersymmetric EFTs as well.
In other words,  we will show that  in many cases supersymmetry allows to relate a  {\it non-trivial} calculation to a
{\it trivial} one (that of the superpartner loops). 
This also provides a way to understand the   few exceptions to the ubiquitous rule  that  $JJ$-operators
 do not renormalize loop-operators at the one-loop level.

The paper is organized as follows.  In Sec.~\ref{U(1)} we start with a simple theory, the EFT  of scalar quantum electrodynamics,
to illustrate our approach  for obtaining one-loop non-renormalization results. In later subsections, we enlarge the theory  including fermions,
and present an exceptional type of $JJ$-operator that  renormalizes loop-operators.   In Sec.~\ref{SM} we show how to generalize our approach to derive analogous results in  the SM EFT and we also discuss the holomorphic properties  of the anomalous dimensions.  In Sec.~\ref{Chiral} we show the implications of our approach for the QCD Chiral Lagrangian. We conclude  in Sec.~\ref{last}.

\section{Non-renormalization results in a $\bma{ U(1)}$  EFT}
 \label{U(1)}
 
 Let us start with the simple  case of a massless  scalar coupled to a $U(1)$-gauge boson with charge $Q_\phi$, assuming for simplicity CP-conservation. 
The corresponding   EFT is defined as an expansion in  derivatives and   fields  over a  heavy new-physics scale 
$\Lambda$:
${\cal L}_{\rm EFT}=\sum_d  {\cal L}_d$,
where ${\cal L}_d$ denotes  the  terms in the expansion made of local operators of dimension $d$.
The leading terms  ($d\leq 6$)   in the EFT are given by
\be
{\cal L}_4=-|D_\mu \phi|^2 -\lambda_\phi |\phi|^4   -\frac{1}{4g^2} F_{\mu\nu}^2\  , \hskip1cm
{\cal L}_6=\frac{1}{\Lambda^2}\left[c_r {\cal O}_r  +c_6{\cal O}_6+c_{FF}{\cal O}_{FF} \right]\, , \label{scalarQED}
\ee
where  the dimension-six operators are 
\be
{\cal O}_r=|\phi|^2 |D_\mu \phi|^2\ ,\hskip.7cm  
{\cal O}_6= |\phi|^6
\ ,\hskip.7cm 
{\cal O}_{FF}= |\phi|^2 F_{\mu\nu}F^{ \mu\nu}\, .
\label{operators6}
\ee
We can use  different bases for the  dimension-six operators although, 
when looking at operator mixing, it is convenient to work in a basis
that separates   $JJ$-operators from loop-operators, as we defined them in the introduction.
Using field redefinitions  (or, equivalently, the equation of motion (EOM) of $\phi$) 
we can reduce the number of  $JJ$-operators to only two: for instance, ${\cal O}_T = \frac{1}{2}J^{\mu} J_{\mu}$ and 
${\cal O}_6= J^* J $, where $J_{\mu}=\phi^* \lra D_\mu\phi$  and $ J=|\phi|^2\phi $.
It is convenient, however, 
to set  a  one-to-one correspondence between operators  and   supersymmetric  $D$-terms, as we will show below.
For this reason, we choose for our basis  ${\cal O}_6$ and ${\cal O}_r$. \footnote{In the $U(1)$ case we are  considering, ${\cal O}_r=\frac{1}{2}\left({\cal O}_H-{\cal O}_T\right)$ where ${\cal O}_H=\frac{1}{2}(\partial_\mu |\phi|^2)^2$.}
The only  loop-operator,  after requiring  CP-invariance, is  ${\cal O}_{FF}$.

Many of the one-loop non-renormalization results that we discuss can be understood from arguments based on the Lorentz structure of the vertices involved. 
Take for instance the non-renormalization of  ${\cal O}_{FF}$ by ${\cal O}_r$.  Integrating by parts and using the EOM, 
we can eliminate  ${\cal O}_r$ in favor of ${\cal O}_r^\prime =  (\phi D_\mu \phi^*)^2 +h.c.$.
Now, it is apparent that ${\cal O}_r^\prime$ cannot renormalize ${\cal O}_{FF}$ because either $\phi D_\mu \phi^*$ or $\phi^* D_\mu \phi$  is external in all   one-loop diagrams, and these Lorentz structures cannot be completed to form  ${\cal O}_{FF}$.  
Since, in addition,  there are no possible one-loop diagrams involving ${\cal O}_6$ that  contribute to  ${\cal O}_{FF}$,
we can conclude that in this EFT   the loop-operator cannot be renormalized  at the one-loop level by the $JJ$-operators.
As we will see, similar Lorentz-based arguments can be used for other non-renormalization results.
This approach, however, requires a case by case analysis and it is not always guaranteed that one can  find an easy argument to see that the loop is zero without  a calculation. 
In this paper  we present a more systematic and unified understanding of such vanishing anomalous dimensions based on a superfield approach that we explain next.

We first  promote the model of \eq{scalarQED} to an ESFT
and study the   renormalization of the dimension-six operators in  this supersymmetric theory.
The superfield formalism makes it transparent to determine  which operators do not  mix at the one-loop level. 
Although in this theory the renormalization of operators involves also loops of  superpartners, we will show in a second step that 
either the ordinary loop  (involving $\phi$ and $A_\mu$) is already trivially zero or it is the superpartner  loops which trivially vanish. 
Therefore, having ensured  that there are no  cancellations between loops of ordinary matter and supermatter, we are able  to extend the supersymmetric non-renormalization results to the non-supersymmetric case. 
In other words, the advantage of this approach is that  we can turn a loop calculation  with the ordinary $\phi$ and $A_\mu$
into a calculation with superpartners,  where the  Lorentz structure of the vertex can make it easier to see that the one-loop contributions are zero.

The dimension-six  operators of \eq{operators6} can be  embedded in different types of  super-operators.  As it will become clear in what follows, it is important for our purposes to embed the dimension-six operators into super-operators with the lowest possible dimension.  This corresponds to an embedding  into  the highest $\theta$-component
of the super-operator (notice that we can always lower the $\theta$-component by adding derivatives in superspace).
This provides a classification of the dimension-six operators that is extremely useful in analyzing the one-loop mixings.
  Let us start with the loop-operator ${\cal O}_{FF}$. 
  Promoting $\phi$ to a chiral supermultiplet $\Phi$ and the gauge boson $A_\mu$  to a  vector supermultiplet $V$, one finds that   ${\cal O}_{FF}$ can be embedded into the $\theta^2$-component ($F$-term)  of the super-operator
\be
\Phi^\dagger  e^{V_\Phi} \Phi \, {\cal W}^\alpha{\cal W}_\alpha=-\frac{1}{2}\theta^2  {\cal O}_{FF}+\cdots \, ,
 \label{nsupersymmetryop}
\ee
where we have defined $V_\Phi\equiv 2Q_\phi V$, ${\cal W}^\alpha$ is the field-strength supermultiplet, and we follow the notation of \cite{Martin:1997ns} (using a mostly-plus metric).
Since the   super-operator in \eq{nsupersymmetryop} is non-chiral,   the ${\cal O}_{FF}$  cannot be generated in a supersymmetry-preserving theory at any loop order.
For the  embedding of the $JJ$-operators, the situation is different.  Some of them can be embedded  in   a $D$-term 
(a $\bar\theta^2\theta^2$-component), while for others this is not possible.
In  the example discussed here, we have
 \be
\left( \Phi^\dagger  e^{V_\Phi} \Phi\right)^2=-4 \theta^2\bar \theta^2{\cal O}_{r}+\cdots\, ,
\ee
and therefore ${\cal O}_{r}$ is allowed by supersymmetry to appear in the K\"ahler potential and is not-protected from one-loop corrections.
Nevertheless ${\cal O}_6$ must arise from the $\theta^0$-component of the super-operator
 \be
\left( \Phi^\dagger  e^{V_\Phi} \Phi\right)^3= {\cal O}_6+\cdots\, ,
\ee
and then must be zero in a supersymmetry-preserving theory at any loop order.

We can now  embed \eq{scalarQED} in a  ESFT.
We use  a  supersymmetry-breaking (SSB) spurion superfield $\eta\equiv \theta^2$ (of dimension $[\eta]=-1$) to
 incorporate the couplings of   \eq{scalarQED}  that break supersymmetry.
We have
 \footnote{Anomaly cancelation requires the inclusion of additional fields that do not play any role in our discussion. We ignore them in what follows.\label{anomalies}}
\bea
{\cal L}_4&\subset &\int d^4\theta \left[\Phi^\dagger  e^{V_\Phi} \Phi+ \lambda_\phi\eta\eta^\dagger (\Phi^\dagger  e^{V_\Phi} \Phi)^2\right]
+\left[\int d^2\theta\, {\cal W}^\alpha {\cal W}_\alpha+h.c.\right]\, ,\nonumber\\
 {\cal L}_6&\subset &\frac{1}{\Lambda^2}
 \int d^4\theta \left\{\tilde c_r\left(\Phi^\dagger  e^{V_\Phi} \Phi\right)^2+\tilde c_6\, \eta\eta^\dagger (\Phi^\dagger  e^{V_\Phi} \Phi)^3
 + \left[ \tilde c_{FF}\,  \eta^\dagger (\Phi^\dagger  e^{V_\Phi} \Phi){\cal W}^\alpha {\cal W}_\alpha+h.c.\right]\right\} .\;\;
\label{scalarSQED}
\eea
It is very  easy to study  the one-loop mixing of the dimension-six operators  in  the above  ESFT  
 using a simple $\eta$-spurion analysis.
 For example, it is clear that there cannot be renormalization  from terms with no SSB spurions, such as $\tilde c_{r}$,
 to terms with SSB spurions, such as $\tilde c_{FF}$.
 Also,  corrections from $\tilde c_{r}$ to $\tilde c_{6}$ are only possible
through the    insertion  of $\lambda_\phi$, that carries a  $\eta\eta^\dagger$.
Similarly,  terms with a SSB spurion $\eta^\dagger$ cannot renormalize terms with two SSB spurions $\eta^\dagger\eta$,
unless they are proportional to $\lambda_\phi$.
This means that $\tilde c_{FF}$  can only  renormalize $\tilde c_{6}$ with the  insertion  of a $\lambda_\phi$.
The inverse is however not guaranteed:
terms with more SSB spurions can in principle renormalize terms with less spurions.
For example,  $\tilde c_{FF}$,  that carries a spurion $\eta^\dagger$, could generate at the loop level the operator
\be
 \int d^4\theta \eta^\dagger \bar {\cal D}^2 {\cal \tilde O}_r=\int d^4\theta (\bar {\cal D}^2\eta^\dagger)  {\cal \tilde O}_r=
\int d^4\theta   {\cal \tilde O}_r\, ,
\ee
where  ${\cal \tilde O}_r=\left(\Phi^\dagger  e^{V_\Phi} \Phi\right)^2$
and  we have defined 
${\cal D}^2\equiv {\cal D}_\alpha{\cal D}^\alpha$, with ${\cal D}_\alpha\Phi=e^{-V_\Phi} D_\alpha (e^{V_\Phi}\Phi)$ 
being  the gauge-covariant derivative in superspace. 
Therefore one has to check it case by case.
For example,  $\tilde c_{6}$ could  in principle  renormalize $\tilde c_{FF}$, 
but  it  is not possible to write the relevant diagram since it involves a vertex with too many $\Phi$'s.
This implies that $\tilde c_{FF}$ is only renormalized by itself at the one-loop level.

This simple renormalization structure is the starting point from which, by examining more closely the loops involved at the field-component  level, we will derive the following non-renormalization results in the non-supersymmetric EFT of \eq{scalarQED}:

{\bf Non-renormalization of  ${\bma {\cal O}_{FF}}$ by  $\bma{\cal O}_{r}$:}
The differences between our original EFT in  \eq{scalarQED} and its supersymmetric version, \eq{scalarSQED}, 
are the presence of the fermion superpartners for the gauge and scalar: the  gaugino, $\lambda$, and  "Higgsino", $\psi$.
We will show, however, that the contributions from superpartners trivially vanish
 in the  mixing of $JJ$- and loop-operators.
In
\be
\int d^4\theta \left( \Phi^\dagger  e^{V_\Phi} \Phi\right)^2=-4{\cal O}_{r}
+2(i \phi^* \lra D_\mu \phi) \psi^\dagger \bsig^\mu\psi+ 2|\phi|^2 (i \psi^\dagger \bsig^\mu \lra{D}_\mu \psi ) +\cdots\, ,
\label{dterm}
\ee
we  have only  the 3 terms shown that can potentially contribute to ${\cal O}_{FF}$ at the one-loop level.
These terms can be considered as part of a supersymmetric $JJ$-operator generated from integrating-out a heavy vector superfield that  contains a scalar, a vector and a fermion. Other terms not shown in \eq{dterm} involve too many fields  (see Appendix) and therefore are only relevant for an analysis beyond one-loop.
The first term of \eq{dterm} can potentially give a contribution to ${\cal O}_{FF}$ from a loop of $\phi$'s,
while the second and third  term could from a loop of Higgsinos.
It is very easy to see that the loop of Higgsinos does not contribute to 
${\cal O}_{FF}$. Indeed,  if in the second term of \eq{dterm}
we close the  Higgsinos  in a loop, the   current $J_\mu=i\phi^* \lra D_\mu \phi$ is left as an external factor, 
and it is then clear that    we can only generate  the  $JJ$-operator $J_\mu J^\mu$. Moreover, the third term of \eq{dterm}  vanishes by using  the EOM: $\bsig^\mu D_\mu\psi =0$ (up to gaugino terms that are not relevant here).
Therefore, Higgsinos do  not  contribute at the  one-loop  level  to the renormalization of the loop-operator ${\cal O}_{FF}$. 
We can then extend the non-renormalization result from the  ESFT of \eq{scalarSQED}
to the non-supersymmetric  EFT of  \eq{scalarQED} and   conclude  that 
 {\it  the loop-operator cannot be renormalized  at the one-loop level by the $JJ$-operators.}

{\bf Non-renormalization of   $\bma{\cal O}_{r}$ by ${\bma {\cal O}_{FF}}$:}
It remains to study  the renormalization from  ${\cal O}_{FF}$  to ${\cal O}_{r}$.
This can arise in principle from a loop of gauge bosons. 
In the supersymmetric theory, \eq{scalarSQED},  $\tilde c_{r}$ does  not carry any SSB spurion and therefore
its renormalization by $\tilde c_{FF}$ cannot be prevented on general grounds, as we explained before.
Nevertheless,  we find that 
operators induced by $\tilde c_{FF}$, through a loop of $V$'s,
must  leave an external factor  $\eta^\dagger\Phi^\dagger  e^{V_\Phi} \Phi$ from the vertex  and  then,  the only  operator that could potentially contribute to  $\tilde c_r$  must have the form
\footnote{Notice that the presence of $\eta^\dagger$, arising  from the vertex,  requires  that the  super-operator must have  two derivatives $\bar  {\cal D}$ in order to potentially contain ${\cal O}_{r}$.}
 \be
\frac{1}{\Lambda^2}\int d^4\theta\,    \eta^\dagger  \left(\Phi^\dagger  e^{V_\Phi} \Phi \right)\bar {\cal D}^2 \left(\Phi^\dagger  e^{V_\Phi} \Phi \right)+h.c.\, .
\label{operator}
\ee
From the  EOM for $\Phi$,  we have that $\bar {\cal D}^2 \Phi^\dagger =0$
up to  $\lambda_\phi$ terms that  bring too many 
powers of $\Phi$, so that the projection of \eq{operator} into ${\cal O}_{r}$ vanishes. 
 Finally, one also has to ensure that 
  redundant $JJ$-super-operators,
that  can give 
$\left(\Phi^\dagger  e^{V_\Phi} \Phi \right)^2$ through superfield redefinitions, are not generated at the one-loop level.
In particular, the redundant super-operator 
\be
\frac{1}{\Lambda^2} \int d^4\theta\,    \left(\Phi^\dagger  e^{V_\Phi} \Phi \right){\cal D}_\alpha {\cal W}^\alpha\, , 
 \label{extraope}
\ee
if generated at the loop level, can give a contribution to $\tilde c_r$ after superfield redefinitions, or equivalently,
after using the  EOM of $V$: ${\cal D}_\alpha {\cal W}^\alpha+h.c.=-gQ_\phi \Phi^\dagger  e^{V_\Phi} \Phi$.
We do not find, however, any non-zero contribution from 
$ \eta^\dagger (\Phi^\dagger  e^{V_\Phi} \Phi) {\cal W}^\alpha{\cal W}_\alpha$  to the operator in \eq{extraope},  as such contributions,  coming from a $V/\Phi$ loop,  
 must  be proportional  to $\eta^\dagger {\cal W}^\alpha \Phi$.~\footnote{Of these, the only one that  cannot be put to zero by the   EOM of $\Phi$ 
 is  $\int d^4\theta\,    \eta^\dagger   {\cal W}^\alpha\Phi[\bar{\cal D}_{\dal},\{{\cal D}_\alpha,\bar{\cal D}^{\dal}\}]e^{V_\Phi}\Phi^\dagger$
but, from the identity $[\bar{\cal D}_{\dal},\{{\cal D}_\alpha,\bar{\cal D}^{\dal}\}]\sim i{\cal W}_\alpha$ \cite{Gates:1983nr}
, one can see that this
 only contributes to $\tilde c_{FF}$.}

Having  shown that  supersymmetry  guarantees  zero contributions to  $\tilde  c_r$ from $\tilde c_{FF}$,
we must  check what are the effects of superpartner loops. From  (see Appendix)
\be
\int d^4\theta \eta^\dagger (\Phi^\dagger e^{V_\Phi} \Phi) {\cal W}^\alpha {\cal W}_\alpha+h.c.
=-{\cal O}_{FF}+\left(\frac{}{}2 i |\phi|^2\  \lambda   \sigma^\mu \partial_\mu \lambda^\dagger
 -\frac{1}{\sqrt{2}}\phi^* \lambda\sigma^{\mu\nu} \psi F_{\mu\nu} 
+h.c.\right)+\dots\, ,
\label{elab}
\ee
where $\sigma^{\mu\nu}= \frac{i}{2} (\sigma^\mu\bar\sigma^\nu-\sigma^\nu\bar\sigma^\mu)$,
it is clear that  a  gaugino/Higgsino loop  cannot  give a contribution to  ${\cal O}_{r}$: the second  term of \eq{elab}, after using the EOM for the gaugino, $\sigma^\mu \partial_\mu \lambda^\dagger=g\phi \psi^\dagger$,  can only give a contribution proportional to  $|\phi|^2\phi$; while the contribution from the  third term must be proportional to  $\phi^*F_{\mu\nu}$. None of them have the right Lorentz structure to contribute to ${\cal O}_{r}$.
Therefore, we conclude that the {\it loop-operator  ${\cal O}_{FF}$ can only renormalize at the one-loop level the 
$JJ$-operators that break supersymmetry, like ${\cal O}_6$, and not 
those that can be embedded in a $D$-term, like ${\cal O}_r$.}

 \subsection{Including fermions}

Let us extend the previous EFT to include  two charged Weyl fermions, $q$ and $u$,  with $U(1)$-charges $Q_q$ and $Q_u$,
such that $Q_\phi+Q_q+Q_u=0$. 
We have now extra terms in the Lagrangian (respecting CP-invariance):
\footnote{Similar remarks to those made in footnote \ref{anomalies} about anomalies apply to this extended model.}
\bea
\Delta {\cal L}_4&=&i q^\dagger \bar\sigma^\mu D_\mu q+ i u^\dagger \bar\sigma^\mu D_\mu u +y_u\left( \phi qu + h.c.\right)\, , \nonumber\\[0.2cm]
\Delta {\cal L}_6&=&\frac{1}{\Lambda^2}\left[
c_{\phi f} {\cal O}_{\phi f}+c_{4f} {\cal O}_{4f} 
+c_{y_u}\left({\cal O}_{y_u}+h.c.\right)
+ c_{D}\left({\cal O}_{D}+h.c.\right) \right]\, , 
\label{scalarQEDextra}
\eea
where $f=q,u$. The $JJ$-operators are
\be
{\cal O}_{y_u} =  |\phi|^2  \phi qu   \, , \ \ \ 
{\cal O}_{\phi f} = i(\phi^* f^\dagger)\bar \sigma^\mu D_\mu (f \phi) \, , \ \ \ 
{\cal O}_{4f} =(f^\dagger\bar\sigma_\mu f)(f^\dagger\bar\sigma^\mu f)\, . \ \ \ 
  \label{extraJJ}
  \ee
Instead of $ {\cal O}_{\phi f}$, we could  have chosen the more  common $JJ$-operator
$ i(\phi^* \lra D_\mu \phi) (f^\dagger\bar \sigma^\mu  f )$  for our basis. Both are related by
\be
\label{Ophifext}
 {\cal O}_{\phi f} = \frac{i}{2}(\phi^* \lra D_\mu \phi) (f^\dagger\bar \sigma^\mu  f )+ \frac{i}{2}|\phi|^2 f^\dagger\bar \sigma^\mu \lra D_\mu f\, , 
\ee
where the last term could be eliminated by the use of the EOM. Our motivation for keeping $ {\cal O}_{\phi f}$ in our basis 
is that, as we will see later,  it is in one-to-one correspondence with a supersymmetric $D$-term.
The  only additional  loop-operator for a $U(1)$ model with fermions is the dipole operator 
\be
{\cal O}_{D} = \ \phi (q \sigma^{\mu\nu} u) F_{\mu\nu} \,  .
\label{dipole}
\ee

\begin{figure}[t]
$$\includegraphics[width=0.25\textwidth]{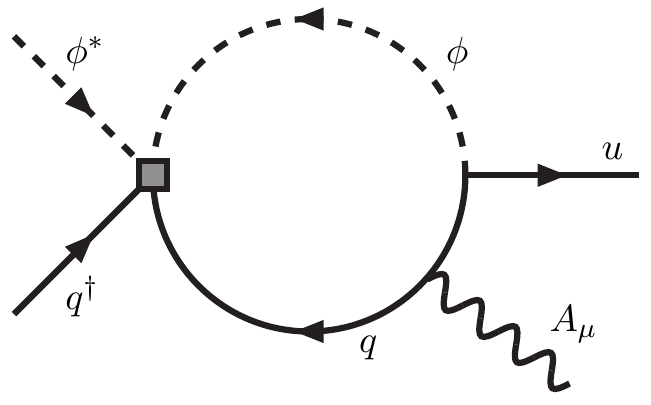} $$
\begin{center}
{
\caption{\emph{A potential contribution from ${\cal O}_{\phi q}$ to ${\cal O}_D$. 
\label{feyn1}}
}}
 \end{center}
\end{figure}

Let us consider the operator mixing in this extended EFT. We will discuss all cases except those for which 
no diagram exists at the one-loop level.  As we said before, in principle, many vanishing entries of the anomalous-dimensions can be simply  understood from inspection of  the Lorentz structure of the different vertices.
For example, it is relatively simple to check that the $JJ$-operators ${\cal O}_{4f}$ and ${\cal O}_{\phi f}$ do not renormalize the loop-operators.    For this purpose, it is  important   to recall that we can write   four-fermion operators, such as  
 $(q^\dagger\bar\sigma_\mu q)(u^\dagger\bar\sigma^\mu u)$, in the equivalent form  $q^\dagger u^\dagger qu$. 
From this, it is obvious that closing a loop of fermions can only give operators  containing  the Lorentz structure
$f^\dagger  f$ or $qu$ that cannot be  completed to give a dipole operator (nor  its equivalent forms, $q  \sigma_{\mu\nu} \sigma_\rho D^\rho q^\dagger F^{\mu\nu}$ or $D_\mu \phi q D^\mu u H$).
  For the case of ${\cal O}_{\phi f}$, the absence of renormalization of the dipole operator, as for example from diagrams like the one in    Fig.~\ref{feyn1},   can be proved just   by realizing  that we can always  keep the  Lorentz structure $\bar\sigma^\mu D_\mu (\phi f)$    external to the loop;   this Lorentz structure cannot be completed to form a dipole operator. The contribution of ${\cal O}_{\phi f}$ to ${\cal O}_{FF}$ is also absent, as can be deduced from  Eq.~(\ref{Ophifext}):
the first term, after  closing the fermion loop,  gives the wrong Lorentz structure to generate ${\cal O}_{FF}$, while
 the second  term   gives an  interaction with too many fields if we use  the fermion EOM.
Finally, ${\cal O}_{y_u}$ can only contribute to the Lorentz structure $\phi qu$, not to the dipole one in \eq{dipole}.

We can be more systematic and complete using our   ESFT approach.
Let us see first how the  operators of \eq{scalarQEDextra} can be embedded in  super-operators. By embedding  $q$ and $u$
in the chiral supermultiplets $Q$ and $U$,
 we find that the dipole loop-operator must  arise from  the  $\theta^2$-term of a non-chiral superfield:
\be\label{susydipole}
\Phi\,  (Q\lra {\cal  D}_\alpha U)\,  {\cal W}^\alpha =-\theta^2  {\cal O}_{D}+\cdots\, .
\ee
Among the  $JJ$-operators of \eq{extraJJ}, two of them can arise from  supersymmetric $D$-terms and are then supersymmetry-preserving:
\be
\left(\Phi^\dagger e^{V_\Phi}\Phi\right)\left(Q^\dagger e^{V_Q}Q\right)=  \bar \theta^2\theta^2 {\cal O}_{\phi q}+\cdots\ ,\ \ \ 
\left(Q^\dagger e^{V_Q} Q\right)\left(Q^\dagger e^{V_Q}Q\right)=-\frac{1}{2}\bar \theta^2\theta^2 {\cal O}_{4q}+\cdots\, ,
\ee
and  similar  operators for    $Q\rightarrow U$,
where we again use the short-hand notation ${V_Q}=2Q_qV$.
 Nevertheless, one of the $JJ$-operators  must come from the $\theta^2$-component of a non-chiral superfield that is not invariant under supersymmetry:
\be
\left(\Phi^\dagger e^{V_\Phi}\Phi\right) \Phi QU=\theta^2 {\cal O}_{y_u}+\cdots\, .
\ee
We can now  promote   \eq{scalarQEDextra} to a ESFT:
\bea
\Delta {\cal L}_4&\subset &\int d^4\theta \left( Q^\dagger  e^{V_Q} Q+U^\dagger  e^{V_U} U\right)+\left[\int d^2\theta\,  y_u \Phi QU+h.c.\right]\, , 
\nonumber \\
\Delta {\cal L}_6&\subset&
\frac{1}{\Lambda^2}\int d^4\theta \Big\{
\tilde c_{\phi f}(\Phi^\dagger e^{V_\Phi}\Phi)(F^\dagger e^{V_F} F)+
\tilde c_{4f}(F^\dagger e^{V_F} F )(F^{\dagger} e^{V_F} F)\nonumber\\
&&+\left[
\eta^\dagger \left(\tilde c_{y_u}  (\Phi^\dagger e^{V_\Phi}\Phi)  \Phi Q U+
\tilde c_D\Phi\, (Q\lra {\cal D}_\alpha U)\,  {\cal W}^\alpha \right)+h.c.\right]\Big\}\, ,
\eea
where $F=Q,U$.

{\bf Non-renormalization of  loop-operators from $\bma{JJ}$-operators:}
The embedding of the EFT into the ESFT shows the  following rule.
Loop-operators  (${\cal O}_{FF}$ and  ${\cal O}_{D}$) cannot be supersymmetrized, while  some $JJ$-operators can be supersymmetrized  (${\cal O}_r$,  ${\cal O}_{4f}$ and  ${\cal O}_{\phi f}$) and others cannot
 (${\cal O}_{y_u}$ and ${\cal O}_6$).
Supersymmetry  then  guarantees that loop-operators can at most  be generated 
from  the latter ones, ${\cal O}_{y_u}$ and ${\cal O}_6$, embedded respectively  in $\eta^\dagger (\Phi^\dagger e^{V_\Phi}\Phi) \Phi QU$ and $\eta\eta^\dagger(\Phi^\dagger e^{V_\Phi}\Phi)^3$.
 By simple inspection of these latter vertices, however, we find that neither of them is possible at
the one-loop level.
  Therefore, in the ESFT the loop-operators are not renormalized at one-loop level by the $JJ$-operators.

To extend  the above results to the  non-supersymmetric EFT, we must
ensure that these non-renormalization results    do not arise from cancellations between loops involving  "ordinary" fields ($A_\mu$, $\phi$, $q$ and $u$) and loops involving superpartners ($\lambda$, $\psi$, $\tilde q$ and $\tilde u$).
This can be  proved  by   showing that either the former  or the latter  are  zero.
In certain cases  it is easier to look at the loop of ordinary fields,  while in others
it is easier to look at the superpartner loops. 
For example,  we have (see appendix)
\be
\int d^4\theta
 \left(Q^\dagger e^{V_Q} Q\right) \left(Q^\dagger e^{V_Q} Q\right)=-\frac{1}{2} {\cal O}_{4q}   +2q^\dagger \bsig^\mu q (i \tq^\dagger \lra D_\mu \tq) +
2(iq^\dagger \bsig^\mu \lra D_\mu  q ) |\tilde q|^2+\cdots\, ,
\label{expansioQ4}
 \ee
where we see that a renormalization to ${\cal O}_{D}$  can arise either from  the first term (by a loop of "quarks" $q$) or the second and third term by a loop of "squarks" $\tilde q$.
It is easier to see that the loops of squarks are zero:  they can only generate operators  containing   $q^\dagger \bsig^\mu q$ or 
$q^\dagger \bsig^\mu \lra D_\mu  q$,  that do not have the  structure necessary to contribute to the dipole operator
${\cal O}_{D}$  nor to
operators related to this one by  EOMs, such as 
$q  \sigma_{\mu\nu} \sigma_\rho D^\rho q^\dagger F^{\mu\nu}$.
 We could proceed similarly for the other operators.
For the case of   ${\cal O}_{\phi f}$, however, 
the one-loop contribution to ${\cal O}_D$ contains  scalars and  fermions (see Fig.~\ref{feyn1}) 
and the corresponding graph with superpartners  has a similar structure, and therefore is not simpler.
Nevertheless, both can be showed to be zero by realizing that 
$\bar\sigma^\mu D_\mu (\phi f)$ can  always be kept as    external to the loop,
and that  this Lorentz structure  cannot be completed to form a dipole operator.
We can conclude that the absence of  renormalization of  loop-operators by $JJ$-operators valid in the ESFT  also
applies to   the EFT.

{\bf Class of $\bma{JJ}$-operators  not renormalized  by loop-operators:}
Following the same  approach, we can also check 
 whether loop-operators can  generate  $JJ$-operators.  
Let us first work within    the ESFT.  
We have shown already that the loop-super-operator 
$\eta^\dagger (\Phi^\dagger  e^{V_\Phi} \Phi){\cal W}^\alpha {\cal W}_\alpha$
cannot generate the $JJ$-super-operator $(\Phi^\dagger  e^{V_\Phi} \Phi)^2$. 
The same  arguments   apply straightforwardly  to $(F^\dagger  e^{V_F} F)(\Phi^\dagger  e^{V_\Phi} \Phi)$.
For the case of  the dipole super-operator, $\eta^\dagger \Phi  (Q\lra {\cal  D}_\alpha U) {\cal W}^\alpha$,
we have a  potential contribution to  $\left(Q^\dagger e^{V_Q} Q\right)\left(U^\dagger e^{V_U} U\right)$ coming from   
a  $\Phi/V$ loop. Nevertheless, as  the factor  $\eta^\dagger Q\lra{\cal D}_\alpha U$  remains  in the external legs,
it is clear that  such contribution can only   lead to  operators  containing  $\eta^\dagger{\cal D}^2$,  which  are not
$JJ$-super-operators. Similarly, contributions to $\left(\Phi^\dagger e^{V_\Phi} \Phi\right)\left(Q^\dagger e^{V_Q} Q\right)$
could arise from a   $U/V$ loop, but one can always arrange it to     leave  either $\eta^\dagger {\cal D}_\alpha \Phi$ or $\eta^\dagger {\cal D}_\alpha Q$ in the external legs
 \footnote{ Using integration by parts and the EOM of $V$, we can write the  dipole super-operator as
 $\int d^4\theta\eta^\dagger \Phi (Q\lra {\cal  D}_\alpha U)\,  {\cal W}^\alpha=-\int d^4\theta \eta^\dagger[({\cal  D}_\alpha \Phi) QU {\cal W}^\alpha+2 \Phi ({\cal  D}_\alpha Q) U {\cal W}^\alpha+O(\Phi_i^5)]$ where $\Phi_i=\Phi,Q,U$.},
 which again does not have the structure of a $JJ$-super-operator (the same applies for $Q\leftrightarrow U$).
Finally we must check  whether redundant  $JJ$-super-operators, as   the one in  \eq{extraope},
can be  generated by  the dipole.  Similar arguments
as those below \eq{extraope} can be used to prove that this is not the case.
Notice, however, that we cannot guarantee the absence of renormalization  by loop-super-operators
 neither of  $\eta^\dagger (\Phi^\dagger e^{V_\Phi}\Phi) \Phi QU$  nor of $\eta\eta^\dagger(\Phi^\dagger e^{V_\Phi}\Phi)^3$.
We then conclude that only the $JJ$-super-operators  that preserve supersymmetry (with no SSB-spurions)
are  safe at the one-loop level from the renormalization by loop-super-operators.

It remains to show that this result extends also to non-supersymmetric EFT.
From \eq{expansionloop} of the Appendix, we have, 
  after using the gaugino EOM and eliminating the auxiliary fields $F_i$,
that loops from superpartners 
can only give contributions  proportional to 
$\phi ff$, $|\phi|^2 f$, $ff$ or $F_{\mu\nu}f$ (for $f=q,u$). None of these terms  can  lead to the Lorentz structure of 
   ${{\cal O}_{r}}$, ${{\cal O}_{4f}}$ nor  ${{\cal O}_{\phi f}}$.  These are exactly  the same $JJ$-operators
that could not be  generated (at one loop) from loop-operators  in the ESFT.

\subsubsection{An exceptional $JJ$-operator}

Let us finally extend the EFT to include an extra fermion, a "down-quark" $d$ of charge  $Q_d$, such that 
$Q_\phi=Q_q+Q_d$.
The following
extra terms are allowed in the Lagrangian:
\bea
\Delta {\cal L}_4&=& id^\dagger \bar\sigma^\mu D_\mu d +y_d\left( \phi^* qd + h.c.\right)\, , \nonumber\\[0.2cm]
\Delta {\cal L}_6&=&
\frac{1}{\Lambda^2}
\left[c_{y_d}{\cal O}_{y_d}+c_{y_uy_d}{\cal O}_{y_uy_d}+h.c.\right] \, , 
\label{scalarQEDextra2}
\eea
where we have the additional $JJ$-operators
\be
{\cal O}_{y_d} =  |\phi|^2 \phi^* qd  \, , \ \ \ 
{\cal O}_{y_uy_d} = quqd \, , \ \ \ 
  \label{extraJJ2}
  \ee
apart from operators similar to  the ones in \eq{scalarQEDextra} with   $f$ including also the $d$. 

Following the ESFT approach, we  embed the $d$-quark
in a chiral supermultiplet $D$   and  the operators of \eq{scalarQEDextra2}  into the super-operators:
\bea
\Phi^\dagger e^{V_\Phi} Q D & = & \theta^2 \phi^* q d + \cdots\ ,\nonumber\\ 
\left(\Phi^\dagger e^{V_\Phi}\Phi\right)\Phi^\dagger e^{V_\Phi}QD&=&\theta^2 {\cal O}_{y_d}+\cdots\ ,\nonumber\\ 
\left(Q U \right) {\cal D}^2\left(Q D \right)&=&  -4 \theta^2{\cal O}_{y_uy_d}+\cdots\, .
\eea
 As all of these operators come from a $\theta^2$-term of non-chiral super-operators,  we learn that  
they can only be generated from  supersymmetry-breaking.
We can promote   \eq{scalarQEDextra2} into a  ESFT in the following way:
\bea
\Delta {\cal L}_4&\subset&\int d^4\theta   \left[D^\dagger e^{V_D} D+\left(\eta^\dagger y_d \Phi^\dagger e^{V_\Phi}QD+h.c.\right)\right]\ ,\nonumber\\
\Delta {\cal L}_6&\subset&
\frac{1}{\Lambda^2}\int d^4\theta\,  \eta^\dagger  \left[
\tilde c_{y_d} \left(\Phi^\dagger e^{V_\Phi}\Phi\right)\Phi^\dagger e^{V_\Phi}QD+
\tilde c_{y_uy_d}\left(Q U \right){\cal D}^2\left(Q D \right)\right]+h.c.\, .
\label{supersymmetry2}
\eea
Now, and this is very important, when considering  only  $d,q,\phi$  in isolation (without the $u$ fermion),  we can always change the supersymmetric embedding of $\phi$ by considering $\phi^*\in \bar \Phi$, where $\bar\Phi$ is a chiral supermultiplet of charge $-1/2$. By doing this, we 
can write the  Yukawa-term for the $d$ in a supersymmetric way, $\int d^2\theta\,  y_d\bar \Phi QD$, and 
guarantee that the  renormalization  of operators involving only  $\phi,q,d$ 
 is identical to the one of    $\phi,q,u$ explained in the previous section. 

It is then clear  that   supersymmetry breaking from Yukawas can only 
arise  through  the combination $y_u y_d$.  
This  allows to explain why  contributions to ${\cal O}_{y_uy_d}$ from
$(q^\dagger\bar\sigma_\mu q)(d^\dagger\bar\sigma^\mu d)$ must be proportional to
$y_u y_d$, as explicit calculations  have  shown in the SM context \cite{Alonso:2014rga}.

\begin{figure}[t]
$$\includegraphics[width=0.25\textwidth]{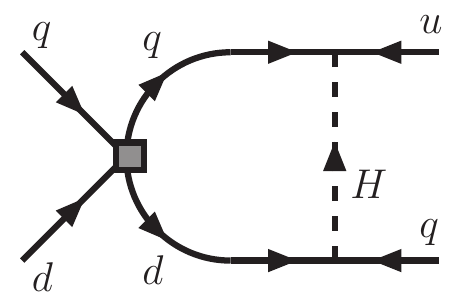}  \quad \quad \quad \quad  \includegraphics[width=0.31\textwidth]{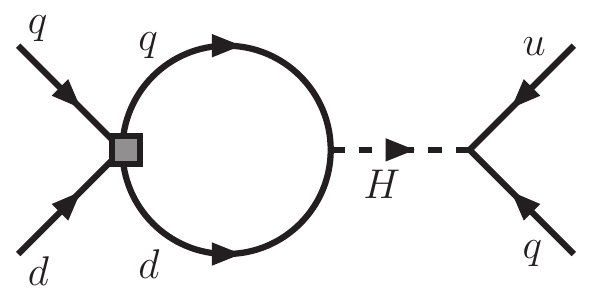} 
$$
\begin{center}
{
\caption{\emph{Contributions to $c_{y_uy_d}$ proportional to $y_d y_u$. }}
\label{feyn2}
}
 \end{center}
\end{figure}

In the ESFT,  the operator $(q^\dagger\bar\sigma_\mu q)(d^\dagger\bar\sigma^\mu d)$
is embedded in a supersymmetry-preserving super-operator
and therefore can only generate supersymmetry-breaking interactions, such as 
${\cal O}_{y_uy_d}$, via  the SSB couplings  $y_u y_d$.
The one-loop contributions from  superpartners do not affect this result, as    \eq{expansioQ4}
shows that they are trivially zero.
 The explicit SM diagrams  are shown in Fig.~\ref{feyn2},
and we have checked that indeed they give a non-zero result.

\vskip.5cm

The operators   ${\cal O}_{y_uy_d}$ and ${\cal O}_{y_{u,d}}$
are the only $JJ$-operators that are embedded in the ESFT with  the same SSB-spurion dependence as the loop-operators -- see \eq{supersymmetry2}.
Therefore, they   can potentially renormalize  ${\cal O}_{D}$. 
Although this was not the case for ${\cal O}_{y_{u,d}}$ due to its Lorentz structure, as we explained above,  we have confirmed by explicit calculation that  ${\cal O}_{y_uy_d}$ indeed renormalizes ${\cal O}_{D}$. This is then an exception to the  ubiquitous rule that $JJ$-operators do not renormalize loop-operators.

\section{Generalization to the Standard Model EFT}
\label{SM}

We can  generalize the previous analysis to dimension-six operators in the SM EFT. 
We begin by constructing an operator basis
that separates $JJ$-operators from loop-operators.
We then classify them  according to their  embedding into a supersymmetric model, depending
on whether they can arise from a super-operator with no SSB spurion ($\eta^0$), which therefore preserves supersymmetry,
or whether they need SSB spurions, either $\bar{\cal D}_{\dot\alpha} \eta^\dagger$, $\eta^\dagger$, $|\bar{\cal D}_{\dot\alpha} \eta^\dagger|^2$ or $\eta \eta^\dagger$ (that selects the $\bar\theta\theta^2$, $\theta^2$,  $\bar\theta\theta$ 
and $\bar\theta^0\theta^0$ component of the super-operator, respectively), or their Hermitian-conjugates. The supersymmetric embedding  naturally selects a SM basis  that we present in Table~\ref{SMclass}.
In this  basis, the non-renormalization results between the different classes of operators discussed in the previous section will also hold.

{\renewcommand{\arraystretch}{1.6} 
\begin{table}[t]  \centering
\begin{tabular}{|c|c||c|c|c|}\cline{1-3} \cline{5-5}  
 \multicolumn{2}{|c||}{Operators} & SSB spurion &    & Super-operators \\ \cline{1-3} \cline{5-5}
\multirow{8}{*}{\rotatebox[origin=c]{90}{$JJ$-operators}} & ${\cal O}_{+}=
D_\mu (H_i^\dagger H_j^\dagger)D^\mu (H^i H^j)
$   &   \multirow{3}{*}{$\eta^0 $}  & \quad \quad   \quad \quad & $(H^\dagger e^{V_H} H)^2$ \\
 & ${\cal O}_{4f}= \left(  \bar f\gamma^\mu t^a f\right)\left(\bar f \gamma_\mu t^a f\right) $  &  & & 
 $( F^\dagger  t^a e^{V_F} F) ( F^\dagger  t^a e^{V_F} F)$ \\   
 & ${\cal O}_{Hf}=
 i (H^\dagger t^a)_i (\bar f t^a)_j \gamma^\mu {D}_\mu  \left( H^i    f^j\right)$  & && 
  $(H^\dagger t^a e^{V_H} H) (F^\dagger t^a e^{V_F} F)$ \\
  \cline{3-3} \cdashline{2-2} \cdashline{5-5}
     & 
 ${\cal O}^{ud}_R= (i H^\dagger \lra D_\mu  \tilde H) (\bar{d}_R \gamma^\mu u_R) $ 
 &   \multirow{1}{*}{ $ \bar{\cal D}_{\dot \alpha}\eta^\dagger$} & & 
$ H^\dagger \bar {\cal D}^{\dot \alpha}\tilde H U^\dagger e^{V_D}D $ 
\\   \cline{3-3} \cdashline{2-2}\cdashline{5-5}
   & ${\cal O}_-=|H^\dagger D_\mu H|^2$  & 
  \multirow{1}{*}{$|\bar {\cal D}_{\dot\alpha}\eta^\dagger |^2$}  & &$|H^\dagger  e^{V_H}{\cal D}_{\alpha}H|^2 $ \\   \cline{3-3} \cdashline{2-2}\cdashline{5-5}
 & ${\cal O}_6= |H|^6$ &  $|\eta|^2$ & & $ (H^\dagger e^{V_H} H)^3 $\\    \cline{3-3}\cdashline{2-2}\cdashline{5-5}
 &   ${\cal O}_y =  |H|^2 H \bar f_{R} f_{L}$  &  \multirow{5}{*}{$\eta^\dagger $} && $(H^\dagger e^{V_H} H) H FF$ \\  
 & ${\cal O}_{yy}= \left(\bar f_{R} t^a f_{L}\right)\left(\bar f_{R} t^a f_{L}\right) $   &  && $(Ft^aF){\cal D}^2(Ft^aF)$ \\[.25cm]   \cline{1-2}\cdashline{5-5}
\multirow{3}{*}{  \rotatebox[origin=c]{90}{ Loop-operators \  }} 
  & ${\cal O}_{D}= H^\dagger \bar f_{R} \sigma^{\mu\nu}t^a f_{L}\,  F^a_{\mu\nu}$ &  &&   $H(F t^a\lra {\cal D}_\alpha F)\,  {\cal W}^{a\, \alpha}$ \\[.16cm] 
 &${\cal O}_{FF_+}= H^\dagger t^a t^b   H F^a_{\mu\nu}( F^{b\, \mu\nu}-i \tilde F^{b\,  \mu\nu})$ &  && $ (H^\dagger t^a t^b e^{V_H} H) {\cal W}^{a\,  \alpha} {\cal W}^{b}_{\alpha}$ \\[.16cm] 
   & ${\cal O}_{3F_+}=f^{abc}F^{a\, \nu}_{\mu}F_{\nu}^{b \, \rho}( F_\rho^{c\, \mu}-i\tilde F_\rho^{c\, \mu})$ &  &&
      $\ f^{abc}{\cal D}^{ \beta} {\cal W}^{a\, \alpha} {\cal W}^b_\beta{\cal W}^c_\alpha$ 
   \\[.16cm] \cline{1-3}  \cline{5-5}
\end{tabular}
\caption{\small {\it Left: Basis of  dimension-six SM operators classified as $JJ$-operators and loop-operators.
We also distinguish those that can arise from a supersymmetric $D$-term  ($\eta^0$) from
those that break supersymmetry either by an spurion $\bar{\cal D}_{\dot\alpha} \eta^\dagger$, $\eta^\dagger$, 
$|\bar{\cal D}_{\dot\alpha} \eta^\dagger|^2$ or $|\eta|^2$. 
We denote by  $F_{\mu\nu}^a$  ($\tilde F_{\mu\nu}^a$) any SM gauge (dual) field-strength. 
The  $t^a$ matrices include   the $U(1)_Y$,   $SU(2)_L$ and $SU(3)_c$ generators, depending on the quantum numbers of the fields involved. 
Fermion operators are written schematically with $f=\{Q_L,u_R,d_R, L_L,e_R\}$. Right: For each operator
in the left column, we provide the super-operator at which it is embedded.}}
    \label{SMclass}
\end{table} 
}
\noindent
The operator basis of Table~\ref{SMclass} is close to 
the basis defined in  Ref.~\cite{Grzadkowski:2010es}.
One significant difference is  our choice of  the  only-Higgs $JJ$-operators, that we take to be  ${\cal O}_{\pm}$
and ${\cal O}_{6}$, and of the Higgs-fermion $JJ$-operator ${\cal O}_{Hf}$. As in the $U(1)$ case, this choice is motivated by the embedding of operators into super-field operators, as we have just mentioned (see more details below). Concerning the classification of 4-fermion operators, our ${\cal O}_{4f}$ operators correspond not only to types $(\bar L L)(\bar L L)$, $(\bar R R)(\bar R R)$ and $(\bar L L)(\bar R R)$ of Ref.~\cite{Grzadkowski:2010es}, but also to
the operator $Q_{ledq}=(\bar L_L e_R)(\bar d_R Q_L)$ classified as $(\bar L R)(\bar R L)$ in \cite{Grzadkowski:2010es},
since this latter can be written  as a ${\cal O}_{4f}$  by Fierz rearrangement.
Finally, our  ${\cal O}_{yy}$ operators correspond to the four  operators of type $(\bar L R)(\bar L R)$ in \cite{Grzadkowski:2010es}.

To embed the SM fields in supermultiplets we follow the common practice of working with left-handed fermion fields so that $Q_L$,
$u^c_R$ and $d^c_R$ are embedded into the chiral supermultiplets $Q$, $U$
and $D$ (generically denoted by $F$). With an abuse of notation,
 we use $H$ for the SM Higgs doublet as well as for the chiral
supermultiplet into which it is embedded. Finally,  gauge bosons are embedded in vector superfields, $V^a$,
and we use the notation $V_\Phi\equiv 2 t^aV^a$
where   $t^a$   include   the  generators of the SM gauge-group in the representation of  the chiral-superfield $\Phi$.

Concerning the embedding of operators into super-operators, there are a few differences  with respect to the $U(1)$ model discussed in the previous section, as we discuss below. Starting with the $JJ$-operators,
we have a new  type of operator not present in the $U(1)$ case,
$ {\cal O}_R^{ud}=  (i H^\dagger \lra D_\mu \tilde  H) (\bar{d}_R \gamma^\mu u_R)$, where $\tilde H \equiv i\sigma_2 H^* $. This operator  cannot be embedded as the others in a $D$-term due to  $\tilde H^\dagger   H=0$ and must  be embedded as  a $ \theta^2\bar \theta $ term of a spinor  super-operator:
\be 
 \int d^4\theta \  \bar{\cal D}_{\dot\alpha}\eta^\dagger( H^\dagger \bar{\cal D}^{\dot\alpha}  \tilde H )  U^\dagger  e^{V_D} D = {\cal O}_R^{ud}+\cdots \, .
\ee 
For the  $JJ$-operators involving only the Higgs field, there is also an important difference with respect to the $U(1)$ case.
We have now two independent operators,~\footnote{The $U(1)$-case identity ${\cal O}_r=({\cal O}_H-{\cal O}_T)/2$ does not hold
 in the SM due to the fact that $H$ is a doublet.}  
 but only one can arise from 
a supersymmetric $D$-term:
\footnote{The operator $(H^\dagger \sigma^a e^{V_H} H)^2$ can be reduced to  $(H^\dagger e^{V_H} H)^2$ by using $\sigma_{ij}^a\sigma^a_{kl}=2\delta_{il}\delta_{kj}
-\delta_{ij}\delta_{kl}$.}
\be
 (H^\dagger e^{V_H} H)^2 = -\bar \theta^2\theta^2 {\cal O}_+ + \cdots\, ,
\label{dos}
\ee
where
\be
\label{JJD}
{\cal O}_{+} = \left[2{\cal O}_r + {\cal O}_H -  {\cal O}_T\right] =
D_\mu (H_i^{\dagger} H_j^{\dagger})D^\mu (H^i H^j)\ ,
\ee
with ${\cal O}_r$,  ${\cal O}_H$ and ${\cal O}_T$
being the SM analogues of the $U(1)$ operators, obtained simply by replacing $\phi$ by $H$. 
The  other independent only-Higgs operator  must arise from a SSB term. We find
that this can be the $\theta \bar\theta$-component of the  superfield 
\be
\bar {\cal D}^{\dot \alpha} ( H^\dagger e^{V_H} H )  {\cal D}_{ \alpha}  ( H^\dagger e^{V_H}H ) 
 = -4 (\bar \sigma^\mu\theta)^{\dot\alpha} ( \sigma^\nu  \bar \theta)_\alpha  \left(  D_\mu H^\dagger H \right)\left( H^\dagger  D_\nu H\right)+\cdots  \, .
\ee
We can write  this operator in a  superfield Lagrangian by using the spurion $|\bar {\cal D}_{\dot\alpha}\eta^\dagger |^2$:
\bea
&&\int d^4\theta\, \ 
\bar {\cal D}_{\dot\alpha}\eta^\dagger  {\cal D}^{\alpha}\eta\  \bar {\cal D}^{\dot \alpha} ( H^\dagger e^{V_H} H )  {\cal D}_{ \alpha}  ( H^\dagger e^{V_H}H ) 
= -16   \ {\cal O}_- +\cdots   \, ,
\label{JJF}
\eea
where
\be
{\cal O}_- =\frac{1}{2}\left[ {\cal O}_H - {\cal O}_T\right]=|H^\dagger D_{\mu}H|^2\ .
\ee

Concerning loop-operators, we have the new operators ${\cal O}_{3F}=f^{abc}F^{a\, \nu}_{\mu}F_{\nu}^{b \, \rho} F_\rho^{c\, \mu}$ and ${\cal O}_{3\tilde F}=f^{abc}F^{a\, \nu}_{\mu}F_{\nu}^{b \, \rho}\tilde F_\rho^{c\, \mu}$, possible now for the non-Abelian
groups $SU(2)_L$ and $SU(3)_c$, which again can only arise from a $\theta^2$-term:
\be
 f^{abc}{\cal D}^{ \beta} {\cal W}^{a\, \alpha} {\cal W}^b_\beta{\cal W}^c_\alpha=
i\theta^2{\cal O}_{3F_{+}}+\cdots \ ,
\label{3W}
\ee 
where we have defined ${\cal O}_{3F_{\pm}}={\cal O}_{3F}\mp i{\cal O}_{3\tilde{F}}$. 
To contain ${\cal O}_{3F_{+}}$, \eq{3W}  must then
appear in the ESFT multiplying   the SSB-spurion $\eta^\dagger$, as the rest of  loop-operators.

For  the loop-operators   ${\cal O}_{FF}=H^\dagger t^a t^b   H F^a_{\mu\nu} F^{b\, \mu\nu}$ and their CP-violating counterparts,    ${\cal O}_{F\tilde F}=H^\dagger t^a t^b   H F^a_{\mu\nu}\tilde F^{b\, \mu\nu}$, 
we can proceed as above and embed them together   in the super-operators 
  \be
  (H^\dagger t^a t^b e^{V_H} H) {\cal W}^{a\, \alpha} {\cal W}^{b}_{\alpha}
=-\frac{1}{2}\theta^2
{\cal O}_{FF_+}+\cdots \, .
 \label{nsupersymmetryop2}
\ee
where
${\cal O}_{FF_{\pm}}={\cal O}_{FF}\mp i{\cal O}_{F\tilde{F}}$.

\subsection{One-loop operator Mixing}

 It is straightforward to extend the  $U(1)$ analysis of  section~2 to the operators of Table~\ref{SMclass} to show   that, with the exception of ${\cal O}_{yy}$,  the $JJ$-operators do not renormalize the loop-operators.
 The only important differences arise  from   the new type of  $JJ$-operators, ${\cal O}_R^{ud}$ and ${\cal O}_-$. Concerning  ${\cal O}_R^{ud}$, it is very simple to see that  this operator cannot renormalize  loop-operators (from a  loop of quarks  one obtains   operators with  the  Lorentz structure $(i\tilde H^\dagger D_\mu H)$;  while  the  Higgs-loop   gives operators  containing    $ \bar d_R\gamma_\mu u_R$, and none of them  can be loop-operators).
Concerning ${\cal O}_-$,  we only need to worry about  the renormalization of  ${\cal O}_{FF}$. 
This can be studied   directly in the ESFT, as  superpartner contributions from $JJ$-operator to loop-operators
can be  shown  to  trivially  vanish. 
In the ESFT, the operator ${\cal O}_-$  is embedded in a super-operator containing 
the SSB-spurion $|{\cal D}_{\alpha}\eta |^2$. This 
 guarantees the absence of renormalization of  loop-super-operators as these latter contain
 the SSB-spurion $\eta^\dagger$.
Besides this direct contribution, there is an indirect route by which  $ {\cal O}_-$  could renormalize ${\cal O}_{FF}$: by generating  ${\cal O}_{HF}=  i  (D^\mu H)^\dagger t^a (D^\nu H) F^a_{\mu\nu}$ which, 
via integration by parts, can give  ${\cal O}_{FF}$.
The operator ${\cal O}_{HF}$ can come from the super-operator
$\tilde {\cal O}_{HF}=\bar{\cal D}_{\dot\alpha} \eta^\dagger \bar{\cal D}^{\dot\alpha}H^\dagger e^{V_H}  {\cal D}_\alpha H\,  {\cal W}^\alpha$  that  in principle is not protected by a simple SSB-spurion analysis from being generated by 
super-operators  $\propto |{\cal D}_{\alpha}\eta |^2$.
Nevertheless, 
contributions to $\tilde {\cal O}_{HF}$ must come from
 \eq{JJF} with derivatives acting on the  two Higgs superfields external to the loop, 
and due to the derivative contractions,  this  can only give
$\bar {\cal D}_{\dot\alpha}\eta^\dagger  {\cal D}^{\alpha}\eta  \bar {\cal D}^{\dot \alpha}  H^\dagger  {\cal D}_{ \alpha}H
{\cal D}_\beta {\cal W}^\beta$; by the use of the EOM of $V$, however, this gives a $JJ$-super-operator and not  $\tilde {\cal O}_{HF}$.
  
In the SM case, the exceptional   ${\cal O}_{yy}$ operators (than can in principle renormalize the dipole operators) are   (following the notation in \cite{EEMP2})
\bea
{\cal O}_{y_uy_d}&=&(\bar Q_L^r   u_R)\epsilon_{rs}(\bar Q_L^s  d_R)\, ,\nonumber  \\
 {\cal O}_{y_uy_d}^{(8)}&=& (\bar  Q_L^r   T^A u_R)\epsilon_{rs} (\bar Q_L^s  T^A d_R)\, ,\nonumber\\
{\cal O}_{y_uy_e}&=& (\bar  Q_L^r  u_R)\epsilon_{rs} (\bar L_L^s e_R)\, , \nonumber \\
 {\cal O}^{\prime}_{y_uy_e}&=&(\bar  Q_L^{r\, \alpha}   e_R)\epsilon_{rs} (\bar L_L^s u^\alpha_R)\, ,
\label{except}
\eea
where $r,s$ are $SU(2)_L$ indices and  $T^A$  are $SU(3)_c$ generators. 
Although in principle all of these four operators could renormalize the SM dipoles, it is easy to realize that ${\cal O}_{y_uy_e}$ will not: the only possible way of closing a loop ($\bar Q_L u_R$ or $\bar L_L e_R$) does not reproduce the dipole Lorentz structure for the external fermion legs. One concludes that only the three remaining operators in \eq{except} renormalize the SM dipole operators and we have verified this by an
explicit calculation.
These are the only dimension-six  $JJ$-operator of the SM that renormalize   loop-operators.
Some of these exceptions were also pointed out in \cite{Jenkins:2013zja}.
Our analysis completes the list of these exceptions and helps to understand the reason behind them. 
From the  analysis of the $U(1)$ case, we can also explain the presence of $y_uy_d$ in the renormalization of 
${\cal O}_{yy}$ from ${\cal O}_{4f}$  \cite{Alonso:2014rga}.

It is obvious that no  operator other than itself renormalizes ${\cal O}_{3F_+}$:
  no adequate one-loop 1PI diagram can be constructed from other dimension-six 
  operators, since they have too many fermion and/or scalar fields.
Nevertheless ${\cal O}_{3F_{+}}$ can in principle renormalize  $JJ$-operators.
Let us consider, for concreteness,  the case  of ${\cal O}_{3F_+}$  made of  $SU(2)_L$ field-strengths.
SM-loop contributions from  ${\cal O}_{3F_{+}}$ can generate  the $JJ$-operators
 $(D_\nu F^{a\, \mu\nu})^2$ and $J^a_\mu D_\nu F^{a\, \mu\nu}$
(where $J^a_\mu$ is the weak current), and indeed these  contributions have been found to be 
 nonzero by an explicit calculation \cite{EGGM}. By using the EOM, $D_\nu F^{a\, \mu\nu}=gJ^{a\, \mu}$, we can reduce  these two operators  to $(J^a_\mu)^2$. Surprisingly,   one finds that the  total contribution from ${\cal O}_{3F_+}$ to  $(J^a_\mu)^2$ adds up to zero \cite{EGGM,Alonso:2014rga}. We can derive  this result as follows.
From  inspection of \eq{s3w}, one can see that the superpartners cannot  give any one-loop contribution to these $JJ$-operators.
Therefore the result must be the same in the SM EFT as  in  the corresponding ESFT. 
Looking at the Higgs component
of  $(J^a_\mu)^2=(H^\dagger\sigma^a \lra D_\mu H)^2+\cdots$, we see that this  operator  must arise from the ESFT term $\int ({\cal D}^{\alpha}\eta {\cal J}^a_\alpha+h.c.)^2$
where ${\cal J}^a_{\alpha}=H^\dagger\sigma^a {\cal D}_{\alpha} H$. 
This super-operator, however, cannot be generated
from  the super-operator in \eq{3W}, as this latter appears in the ESFT with a different number of SSB-spurions, $\eta^\dagger$. This proves that ${\cal O}_{3F_+}$ cannot generate
$JJ$-operators with  Higgs. Now,   if  current-current super-operators with $H$ are not generated,
 those with $Q$ cannot be generated either, since  in the ESFT the $SU(2)_L$ vector does not distinguish between different  
$SU(2)_L$-doublet chiral superfields. This completes the proof that ${\cal O}_{3F_+}$ does not renormalize any $JJ$-operator in the basis of  Table~\ref{SMclass}.

 Concerning the non-renormalization of $JJ$-operators by loop-operators, the last new case left to discuss is that of ${\cal O}_-$ by ${\cal O}_{FF}$. The SSB-spurion analysis forbids such renormalization in the ESFT and the result can be extended to the SM EFT as no superpartner-loop contributes either (see Eq.~(\ref{expansionOFF}) in the Appendix).

At energies below the electroweak scale, we can  integrate out $W$, $Z$, Higgs  and  top,
and write an EFT with only light quarks and leptons, photon and gluons.
This EFT contains  four-fermion operators of type ${\cal O}_{4f}$, generated at tree-level, that are  $JJ$-operators, and other operators
of dipole-type  that are loop-operators.
Following the  above approach  we can prove that these four-fermion operators cannot renormalize the  dipole-type operators,
and this is exactly  what is found in explicit calculations \cite{Grinstein:1990tj}.

\subsection{Holomorphy of the anomalous dimensions}

\begin{figure}[t]
$$ \includegraphics[width=0.32\textwidth]{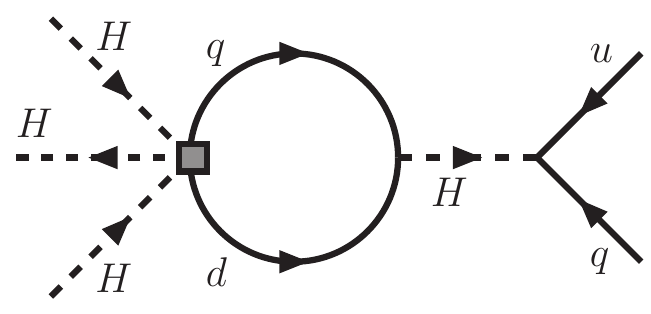} \quad\quad \quad\quad   \includegraphics[width=0.27\textwidth]{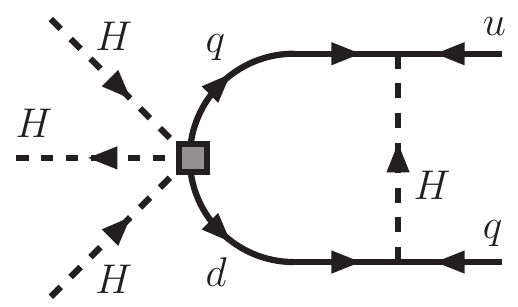} $$
\begin{center}
{
\caption{\emph{   Non-holomorphic mixing between ${\cal O}_{y_u}$ and  ${\cal O}_{y_d}$.
\label{nonholo}}
}}
 \end{center}
\end{figure}

It has been recently shown  in  Ref.~\cite{Alonso:2014rga},  based on  explicit calculations,  that the  anomalous dimension matrix
respects, to a large extent, holomorphy. 
Here we would like to show how to derive some of  these properties  using our ESFT approach.
In particular, we will derive  that, with the exception of one case, the one-loop anomalous dimensions of the complex Wilson-coefficients  $c_i=\{ c_{3F_+}, c_{FF_+},c_D,c_y,c_{yy},c_{R}^{ud}\}$ do not 
depend  on their  complex-conjugates  $c_j^*$:
\be
\frac{\partial \gamma_{c_i}}{\partial c_j^*} = 0 \, .
\label{hcond}
\ee
We  start by showing  when \eq{hcond} is  satisfied just by simple inspection of the SM diagrams.
For example,  it is easy to realize that holomorphy must be  respected in  contributions  from dimension-six operators in which fermions  with a given chirality, e.g., $f_\alpha$ or  $f_\alpha f'_\beta$,   are kept  as external legs; indeed,
 the corresponding Hermitian-conjugate operator can only contribute to  operators with  fermions in the opposite chirality. 
 Interestingly, we can extend the same argument to operators with field-strengths
if we  write the  loop-operators as 
\be
{\cal O}_{3F_+}=
 -\frac{1}{4} \text{tr} \ {\cal F}_{\alpha}^{\ \beta}  {\cal F}_{\beta}^{\ \lambda}   {\cal F}_{\lambda}^{\ \alpha }\, ,
\ \ {\cal O}_{FF_+} = 
\frac{1}{4}  H^\dagger t^a t^b H  ({\cal F}^a)_{\alpha\beta} ({\cal F}^b)^{\beta\alpha}\, ,\  \ 
{\cal O}_{D} = H^\dagger f_\alpha ({\cal F}^a)^{\alpha\beta}t^a f'_\beta  \, ,
\label{explicithol}
\ee
where we have defined ${\cal F}^{\alpha\beta} \equiv (F^a_{\mu\nu}t^a\sigma^{\mu\nu})^{\alpha\beta}$
that  transforms as a  $\bf (1,0)$ under the Lorentz group, and write the Hermitian-conjugate of \eq{explicithol} with  ${\cal F}^{\dot\alpha\dot\beta}$, a $\bf (0,1)$ under the Lorentz group, 
as for example, ${\cal O}^\dagger_{3F^+}  ={\cal O}_{3F^-}  = -\frac{1}{4} \text{tr} \ {\cal F}_{\dot\alpha}^{\ \dot \beta}{\cal F}_{\dot\beta}^{\ \dot \lambda}   {\cal F}_{\dot\lambda}^{\ \dot\alpha }$.
From \eq{explicithol}  it is clear  that any diagram with an external ${\cal F}_{\alpha\beta}$  respects holomorphy,
as it can only generate the operators of \eq{explicithol}  and not their Hermitian conjugates. 
One-loop contributions from ${\cal O}_{FF_+}$ in which  $H^\dagger t^a t^b H$ is kept   among the  external fields, however,
do not necessarily respect holomorphy. 
An explicit calculation is needed, and      
while  contributions  to  ${\cal O}_{FF_+}$ vanish by the  reasoning given in \cite{GJMT},
contributions to ${\cal O}_{y}$  are found to be holomorphic. Furthermore, potentially non-holomorphic contributions can also be  identified by tracking the fermion chirality. For instance, in the   ${\cal O}_{y_u}\leftrightarrow {\cal O}_{y_d}$ operator mixing,  the fermion chirality flips due to contributions proportional to $y_u y_d$, as shown in Fig.~\ref{nonholo}.  
These contributions  have the same loop structure as the contributions of Fig.~\ref{feyn2}.  

Following our previous supersymmetric approach,  it is quite simple to check whether or not loop contributions are holomorphic. 
In the ESFT, holomorphy is trivially respected as super-operators with an $\eta^\dagger$-spurion renormalize among themselves and cannot induce the Hermitian-conjugate 
super-operators since  those  contain an  $\eta$, and vice versa. The only exception to this rule can arise
from  supersymmetry-breaking, which requires  the combination $y_u y_d$. 
We find that the only  possible contributions $\propto y_u y_d$ 
are the supersymmetric versions of the diagrams in Fig.~\ref{nonholo}, that generate, as we said, 
 a mixing between ${\cal O}_{y_u}$  and ${\cal O}_{y_d}$.
 
\begin{figure}[t]
$$\includegraphics[width=0.32\textwidth]{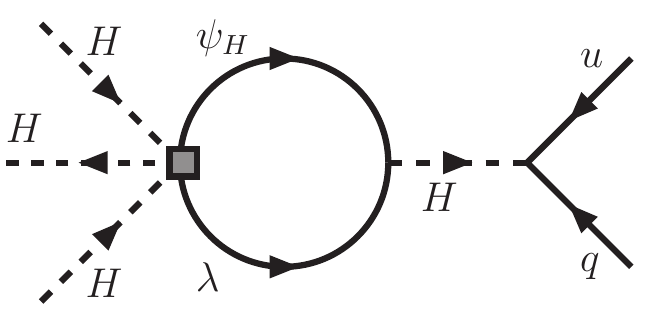} \quad\quad\quad\quad \includegraphics[width=0.28\textwidth]{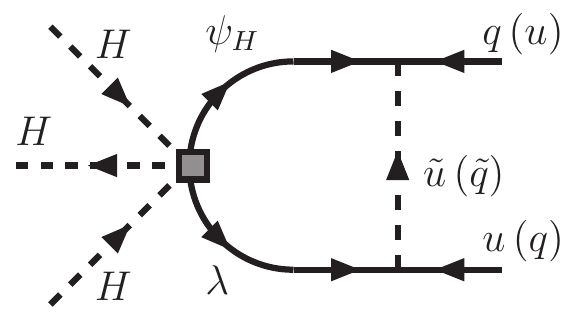} $$
\begin{center}
{
\caption{\emph{ Contributions from   ${\cal O}^\dagger_{FF^+}$ to ${\cal O}_{y}$.
\label{nonholosusy}}
}}
 \end{center}
\end{figure}

Given this holomorphy  property of  the ESFT, we still must check whether  at the field-component level this property   is preserved, 
and assure that it is not due to cancellations between SM contributions and those of superpartners. 
This can be easily done by  looking at either one or the other loop (whenever holomorphy in the ESFT implies that the sum is zero).
In this way, as one of the loops (either the SM  or the superpartner one) 
involves fermions, we can always  relate holomorphy to fermion chirality.
We find that the only potentially non-holomorphic contributions come from the diagrams in Fig.~\ref{nonholosusy}, which   correspond  to the superpartner one-loop contributions to ${\cal O}_y$ arising from  ${\cal O}^\dagger_{FF_+}$ (illustrated for the $U(1)_Y$ case).
These contributions  are induced by the vertex  $|H|^2  \lambda^\dagger   \bar \sigma^\mu \partial_\mu \lambda\sim 
  |H|^2 H  \lambda^\dagger     \psi_H^\dagger$  of \eq{elab}, where we have used the EOM of $\lambda$ (and 
replaced the $U(1)$ $\phi$ and $\psi$ by the SM Higgs and Higgsino).   
An explicit calculation of these diagrams  shows however
 that  they cancel each other, the sum being proportional to $Y_H+Y_q+Y_u=0$. A
similar cancellation occurs in the $SU(2)$ case.  This could have been anticipated as  there are no such diagrams for the  $SU(3)_c$ case,
and this cancellation should be independent of the gauge group.

 We  conclude that  the only non-holomorphic anomalous dimension  is in the ${\cal O}_{y_u}\leftrightarrow {\cal O}_{y_d}$ mixing, and its origin can be tracked  to the supersymmetry-breaking combination $y_u y_d$.

\section{Implications for the QCD Chiral Lagrangian}\label{Chiral}

We can extend the above analysis also to
the QCD Chiral Lagrangian  \cite{Gasser:1983yg}.  
At $O(p^2)$, we have
\be
{\cal L}_2=\frac{f^2_\pi}{4}\langle D_\mu U^\dagger D^\mu  U\rangle\, .
\label{leadingchiral}
\ee
This is an operator that can be embedded in a $D$-term as $\int d^4\theta\, \langle{\cal U}^\dagger {\cal U}\rangle$, where $U$ and its superpartners are contained in ${\cal U}\equiv e^{i\Phi}$, with $\Phi$  being  a chiral superfield.
At $O(p^4)$, the QCD Chiral Lagrangian is usually  parametrized by the $L_i$ coefficients \cite{Gasser:1983yg} in a basis
 with  operators  that are linear combinations of  $JJ$-operators and loop-operators. These are
\be
{\cal L}_{4} = - iL_9 \langle F_R^{\mu\nu}D_\mu U D_\nu U^\dagger + F_L^{\mu\nu} D_\mu U^\dagger D_\nu U\rangle  + L_{10}\langle U^\dagger F_R^{\mu\nu} U F_{L\mu\nu}\rangle+\cdots \, .
\ee
A more convenient basis is however
 \be
{\cal L}_4=i L_{\!J\!J}   \langle  D_\mu F_L^{\mu\nu}(U^\dagger \lra{D}_\nu U)+(U \lra{D}_\nu U^\dagger) D_\mu F_R^{\mu\nu}\rangle
+ L_{loop}\langle U^\dagger F_R^{\mu\nu} U F_{L\mu\nu}\rangle
+\cdots\  ,
\label{chiral}
 \ee 
where $L_{\!J\!J}=L_9/2$ and $L_{loop}=L_{9}+L_{10}$.
It is easy to see that  the first operator of \eq{chiral} is a  $JJ$-operator,
while  the second is a loop-operator.
 This latter  can only be embedded in a $\theta^2$-term of a super-operator (i.e., $\langle{\cal U^\dagger W}^\alpha_R\, {\cal  U\, W}_{\alpha L}\rangle$),
  and therefore it cannot be renormalized by the operator in \eq{leadingchiral}  in the supersymmetric limit.
  As  contributions from  superpartner 
    loops
    can  be easily shown to
     vanish, we can deduce that \eq{leadingchiral} cannot renormalize $L_{loop}$ at the one-loop level. 
This is indeed what one finds from the explicit calculation \cite{Gasser:1983yg}:
$\gamma_{L_{loop}}=\gamma_{L_9}+\gamma_{L_{10}}=1/4-1/4=0$.

\begin{figure}
$$\includegraphics[width=0.65\textwidth]{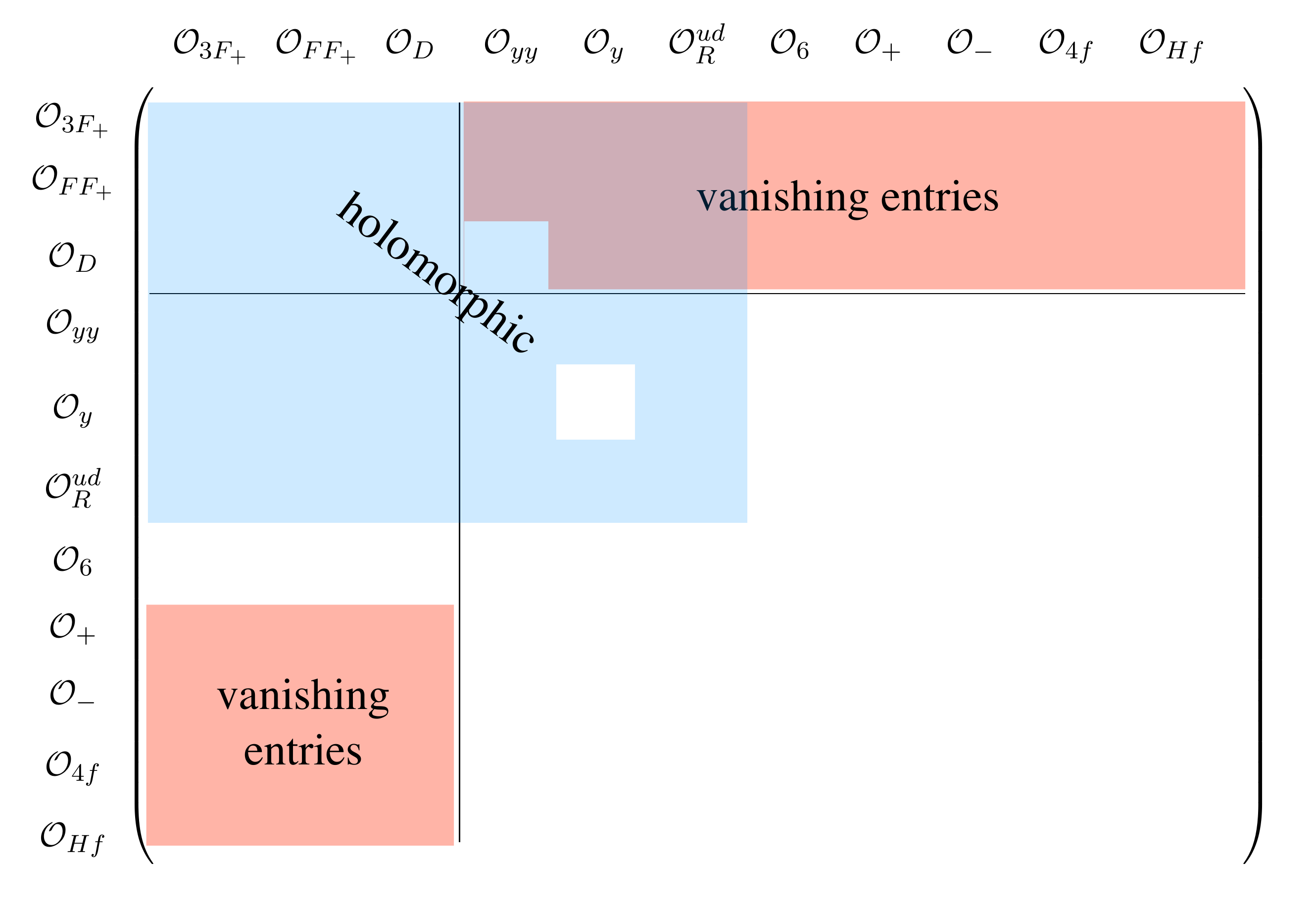} $$
\begin{center}
{
\caption{\emph{Anomalous-dimension matrix of the dimension-six SM operators showing  
which entries (red-shaded)  vanish  following the present analysis.
We also show the   entries (light blue-shaded) that respect the holomorphic condition \eq{hcond}. 
Solid lines separate loop-operators from $JJ$-operators. 
\label{msintesis}}
}}
 \end{center}
\end{figure}

\section{Conclusions}
\label{last}

In EFTs with higher-dimensional operators the 
one-loop anomalous dimension matrix  has plenty of vanishing entries apparently not forbidden by the symmetries of the theory. 
In this paper we have shown that the reason behind these zeros is the different Lorentz structure of the operators
that does not allow them to mix at the one-loop level.
We have proposed a way to  understand the pattern underlying these zeros based on  classifying the dimension-six operators 
in $JJ$- and loop-operators and also
according to  their  embedding in super-operators (see  Table~\ref{SMclass} for the SM EFT).
We have seen that all  loop-operators  break supersymmetry,~\footnote{This is not true in general. For instance, in models with two Higgses of opposite hypercharge, 
$H$ and $\bar H$,
one can have the supersymmetric loop-operator $\int d^2\theta H \bar H {\cal W}^\alpha {\cal W}_\alpha$. Notice that in such a case supersymmetry also protects that operator from being renormalized in the ESFT.} while
 we have two classes of $JJ$-operators, those that   can be    supersymetrized and those that cannot.
This classification  is  very useful   to obtain  non-renormalization results  based 
in  a pure  SSB-spurion analysis  in superfields,
that can be extended to  non-supersymmetric EFTs.
In terms of component fields, the crucial point is that
the vanishing of the anomalous-dimensions  does not arise from  cancellations between bosons and fermions
but from the  underlying Lorentz structure of the operators.

We have presented how this approach works  in a simple $U(1)$ model with a scalar and fermions, 
and have explained  how to extend this to  SM EFTs and  the QCD Chiral Langrangian.
The main results are summarized in Fig.~\ref{msintesis} that   shows which entries of the anomalous-dimension matrix for the SM EFTs operators  we have proved  to vanish. 
We have also explained how to check if holomorphy is respected by the complex Wilson-coefficients,
a property that is fulfilled in most   cases, as Fig.~\ref{msintesis} shows.
Our approach can be generalized to other theories as well
as  to the  analysis of  other  anomalous dimensions, a work that we leave for a further publication.

\section*{Note added}
We have corrected the holomorphic structure of the anomalous dimension matrix  with 
respect to the previous version of this paper (see Fig.~\ref{msintesis}) profitting from the analysis of 
Ref.~\cite{Cheung}.

\section*{Acknowledgments}
We thank Eduard Masso for very useful discussions and collaboration in the early stages of this work,
and Clifford Cheung and Rodrigo Alonso for useful discussions and clarifications.
We acknowledge support by the Spanish Ministry MEC under grants FPA2013-44773-P, FPA2012-32828, and FPA2011-25948, by the Generalitat grant 2014-SGR-1450 and by
the Severo Ochoa excellence program of MINECO (grant SO-2012-0234).
The work of J.E.M. has been supported by the Spanish Ministry MECD through the FPU grant AP2010-3193.
The work of A.P. has also been supported by  the Catalan ICREA Academia Program.

\section*{Appendix}

In this Appendix we  show  the expansion  in component fields
of some of the  super-operators discussed in the text. 
We   work  in the Wess-Zumino gauge. 
In particular,   for the $U(1)$ case,  
we show   the supersymmetry-preserving super-operator
  \bea
&& \int d^4\theta    \left( \Phi^\dagger e^{V_\Phi}  \Phi  \right) \left(Q^\dagger e^{V_Q} Q \right)   = -|\tq|^2 |D_\mu \phi|^2-|\phi|^2 | D_\mu \tq|^2- \frac{1}{2}\partial_\mu |\tq|^2  \partial^\mu |\phi|^2  +\frac{i}{2} |\tq|^2 (\psi^\dagger \bar \sigma^\mu \lra D_\mu \psi )\nonumber\\[0.2cm]
&&\quad   +\frac{i}{2}|\phi|^2( q^\dagger \bar \sigma^\mu \lra D_\mu q )   +\frac{1}{2} \left[ (\psi^\dagger \bar\sigma^\mu q)(i \tq^* \lra D_\mu \phi )+h.c.\right]+\frac{1}{2}\left[\phi\tq^* (i\psi^\dagger \bar\sigma^\mu \lra D_\mu q)+ h.c.\right]
 \nonumber \\[0.2cm]
 &&\quad - \frac{1}{2} \left( i \phi^* \lra{D}_\mu \phi   -\psi^\dagger \bar\sigma^\mu \psi  \right) \left( i \tq^* \lra{D}_\mu \tq   -q^\dagger \bar\sigma^\mu q  \right)    \nonumber\\[0.2cm]
&&\quad  -\left[ (\psi^\dagger q^\dagger)\phi F_q  +(\psi^\dagger q^\dagger)  \tq  F_\phi- \phi F_\phi^* \tq^* F_q  + h.c.\right] + |\phi|^2|F_q|^2 +|\tilde q|^2|F_\phi|^2 \nonumber\\[0.2cm]
 &&\quad-\sqrt{2}g(Q_\phi+Q_q)\left[|\phi|^2\tilde q \lambda^\dagger q^\dagger-|\tq|^2 \phi  \lambda^\dagger \psi^\dagger  +h.c.\right]+ g (Q_\phi+ Q_q ) |\phi|^2|\tq|^2  D 
\, ,
\eea
where boundary terms have been dropped out in integration by parts rearrangements. 
The fields are embedded in the super-multiplets as $\Phi\sim \{\phi, \psi, F_\phi \}$, $Q\sim\{\tilde q, q, F_q \}$ and  $V\sim\{\lambda, A_\mu, D \}$.
The $D$ and $F_{q,\phi}$ auxiliary fields are irrelevant in the discussion of the renormalization of loop-operators by $JJ$-operators because they are necessarily involved in vertices with too many scalar and/or fermion fields.

The loop-super-operators for the $U(1)$ case
are given by
\bea
\label{expansionOFF}
\int d^4\theta \,  \eta^\dagger  (\Phi^\dagger e^{V_\Phi} \Phi) {\cal W}^\alpha {\cal W}_\alpha  &=& -\frac{1}{2}{\cal O}_{FF_+}+ |\phi^2|\left( D^2 + 2 i \lambda \sigma^\mu \partial_\mu  \lambda^\dagger \right)  \nonumber\\
&&-  \frac{1}{\sqrt{2}} \phi^* \lambda\sigma^{\mu\nu}  \psi F_{\mu\nu}-\sqrt{2}\phi^* \psi \lambda D +\lambda\lambda \phi^* F_\phi \, , \\[.2cm]
 \int d^4 \theta\, \eta^\dagger \Phi (Q \lra {\cal D}_\alpha U){\cal W}^\alpha &=&-{\cal O}_D   +  \Big\{ - \sqrt{2} i   \phi  \tilde{q} (u \sigma^\mu     \partial_\mu  \lambda^\dagger )+2\tilde{u}\phi F_q D+2 \sqrt{2} \phi   F_u  \lambda  q \nonumber\\
 &&  +   \sqrt{2}  \tilde{u} F_\phi   \lambda q+ \sqrt{2}F_u \tilde {q} \psi  \lambda+   D \tilde{q} \psi u     \nonumber\\
&&- \frac{1}{2} \tilde{q} \psi \sigma_{\mu\nu} uF^{\mu\nu}  +\sqrt{2}(  \psi q) (\lambda u)    - (u\leftrightarrow q )\Big\}      \, .\ \  \ \ \  \  \ 
 \label{expansionloop}
  \eea
For the non-Abelian case, there is  also the loop-super-operator
\be
\int d^4 \theta \  \eta^\dagger  \text{tr}[{\cal D}^{ \beta} {\cal W}^{ \alpha} {\cal W}_\beta{\cal W}_\alpha]
= \frac{1}{4} {\cal O }_{3F^+} + i\   \text{tr}\left[  \frac{1}{2}   F_{\mu\nu}  \lambda  \sigma^{\mu\nu}( \sigma^\gamma  \partial_\gamma \lambda^\dagger)    +   \     \lambda \sigma^\mu \partial_\mu \lambda^\dagger D \right] \, .
\label{s3w}
\ee



\end{document}